\documentclass[graybox]{svmult}
\usepackage{makeidx}         
\usepackage{graphicx}        
\usepackage{multicol}        
\usepackage[bottom]{footmisc}

\usepackage{newtxtext}       %
\usepackage[varvw]{newtxmath}       

\usepackage[english]{babel}
\usepackage[utf8]{inputenc}
\usepackage[pdftex, pdftitle={Article}, pdfauthor={Author}]{hyperref} 
\begin{document}
\title*{On Stability and Isotropization of Kasner Solution in $R^{2}$ Gravity}
\author{Dmitri Pogosyan \orcidID{0000-0002-7998-6823} and \\
Akash Kav \orcidID{0009-0005-7525-2100}}
\institute{Dmitri Pogosyan \at University of Alberta, Edmonton, Alberta, Canada T6G 2E1, \email{pogosyan@ualberta,ca}
\and Akash Kav \at University of Alberta, Edmonton, Alberta, Canada T6G 2E1,
\email{kav@ualberta.ca}}
\maketitle

\abstract{
The fate of the Universe that initially expands anisotropically in the theory with $R^{2}$ quantum-gravitational term in
the Lagrangian is investigated.  The stability of Kasner-like expansion, specifically in the class of Kantowski-Sachs spacetimes, is analyzed.
Kasner solutions are found to be unstable, with the bifurcation line between the initial conditions that lead to collapsing universes and
the ones that set the universe for continuing expansion that becomes isotropic, established analytically. Under suitable conditions,
the isotropized spacetime enters the intermediate slow-rolling inflationary stage similar to Starobinsky inflation.
}

\section{Introduction} \label{sec:Introduction}

Forty five years ago, A. A. Starobinsky published a seminal paper \cite{Starobinsky} in which he suggested the novel cosmological scenario
with homogeneous and isotropic inflationary expansion of the Universe based on quantum-gravitational corrections to the vacuum energy-momentum tensor of the matter fields \cite{Davies}.  Arguably the first complete and physically motivated model of inflation, Starobinsky scenario 
put what became known as $R^2$ modified gravity at the forefront of cosmological research into the very early Universe for decades that follow.

$R^2$ gravity also admits anisotropic solutions, in particular the exact Kasner \cite{kasner} solutions, that play a central role in
the Belinsky-Khalatnikov-Lifshitz (BKL) theory of generic anisotropic behaviour of cosmological models near singularity\cite{belinsky}
where vacuum effects are dominant.

Shortly thereafter, Starobinsky tasked one of the authors (DP), then an undergraduate student, with the question whether there is a possible transition from
anisotropic expansion of the Universe to isotropic behaviour if quantum-gravitational effects are taken into account.  With the help of 
numerical coding and punch cards,
such possibility was demonstrated,  the term project successfully defended, but the result was never published.  Now it is a turn 
for a new generation of undergraduate students to make their contribution.

The fundamental equations of cosmology are the Einstein field equations (EFEs),  proposed by Albert Einstein in 1916. These equations can be derived from the principle of least action, where a Lagrangian density linear in the Ricci scalar, $R$ is considered. 
\begin{equation}
\delta S = \delta \int \left( \frac{R}{16\pi G} + \mathcal{L}_m \right) \sqrt{-g} d^{4}x = 0
\end{equation}
and take the following form
\begin{equation}
R_{\mu \nu} - \frac{1}{2}g_{\mu \nu}R = -8 \pi G T_{\mu \nu}
\end{equation}
 with $R_{\mu \nu}$ being the Ricci tensor, $g_{\mu \nu}$ the metric tensor with $g$ its determinant, and $T_{\mu \nu}$ the energy-momentum tensor
 arising from the  matter Lagrangian $\mathcal{L}_m$.\footnote{ $c=\hbar=1$ system of units is used hereafter. In these units, the Newton's constant is inverse
 square of the Planck mass $G=1/M_{pl}^2$}
 
In the early universe, quantum fluctuations in the matter fields in vacuum lead to the contribution to the energy-momentum tensor that is quadratic in $R$ \cite{Davies}
\begin{equation}
R_{\mu \nu} - \frac{1}{2}g_{\mu \nu}R = -\kappa T_{\mu \nu}(R^{\alpha \beta}R_{\alpha \beta}, R^{2}, ...)
\end{equation}
It was found by Ginzburg et al. \cite{Ginzburg} that in four dimensions in a general space-time, only two combinations of quadratic curvature terms conserve the energy-momentum tensor . These combinations can be obtained from varying $R^{2}$ and $R_{\mu \nu} R^{\mu \nu}$ terms in the Lagrangian density
\begin{equation}
\mathcal{L} = \sqrt{-g}f(R_{\mu \nu} R^{\mu \nu}, R^{2})
\end{equation}
Although by now theories of ghost-free gravity have been developed in the Einstein frame \cite{Nojiri}, in order to avoid complications with ghost degrees of freedom, exclude the problematic $R_{\mu \nu}R^{\mu \nu}$ from consideration.
Thus, we elect to consider only the case of a Lagrangian density with terms linear and quadratic in $R$.
\begin{equation} \label{eq:LagrangianDensity}
\mathcal{L} = \frac{1}{16\pi G} \left(R + \frac{R^{2}}{m^{2}}\right) ~.
\end{equation}
Here $m$ is a constant parameter, which defines a curvature scale upon which quantum corrections begin to become relevant.  

There can be also another, model dependent contribution to the renormalized energy momentum tensor, that cannot be put as a variation of any Lagrangian density. 
This term is known as the conformal anomaly or trace anomaly. The effect of this term was explored in homogeneous and isotropic geometries first by Fischetti et al. \cite{Fischetti} and its main role is in creating an exact de-Sitter solution for such space-times.  
This solution motivated Starobinsky's original paper, but its the subsequent inflation during slow roll stage after de-Sitter state decay
is what provides a more realistic scenario for the Universe.  This slow roll inflation including its smooth transition to Friedmann stage with dust-like
equation of state is governed by $R^2$ Lagrangian contribution.  It serves as an attractor to the majority of trajectories in the system's phase 
space and therefore describes the generic behaviour of isotropic universes. The complete picture 
was given a detailed treatment by Kofman, Muchanov and Pogosyan in \cite{KMP1987}. 

In this paper we study spatially flat homogeneous but initially anisotropic models of the Universe  with the $R^2$ Lagrangian density  given by Eq.~\ref{eq:LagrangianDensity} and no conformal anomaly.  The metric interval that includes anisotropy 
can be written with three scale factors, $a(t), b(t), c(t)$
\begin{equation}
ds^{2} = dt^{2} - a(t)^{2}dx^{2} - b(t)^{2}dy^{2} - d(t)^{2}dz^{2}
\end{equation}
In the absence of quantum effects, there are no isotropic $a=b=c$ vacuum solutions, and anisotropic Kasner power law solutions \cite{kasner} are generic.
With quantum correction present, homogeneous solutions are possible, and while Kasner solutions still remain the exact ones, how stable and
general they are is the subject of this paper's investigation.

We limit our study to the class of Kantowski-Sachs spacetimes \cite{Kantowski}, which can be thought of as a special case of the Kasner metric, where two scale factors out of three are equal.  After brief review of isotropic case in Sect.~\ref{sec:Quadratic EH}, we derive equations for $R^{2}$ gravity with an anisotropic metric in Sect.~\ref{sec:AnisotropicGravity}, and explore the solutions with parameters describing the anisotropy and volume for an initially expanding universe.

Noting that the Kantowski-Sachs metric is an exact solution to the EFEs in $f(R)$ gravity, as was shown, for example, by Shamir \cite{Shamir}, we recognize that a solution which begins on the Kasner solutions exactly will remain there and not tend away from it. Thus, similar to the work of Toporensky and Müller \cite{Toporensky}, as well as Barrow and Hevrik \cite{Barrow}, we analyze the stability of these Kasner solutions by considering a perturbation off these solutions. The work of Müller and Toporensky considered the case tending toward the BKL singularity \cite{belinsky}, approaching which the universe oscillates between different directions that collapse or expand. We consider the case tending away from the BKL singularity, when the comoving volume of space initially grows, and look to find a bifurcation line between the regimes which develop, as well an effect of isotropization, by evaluating the equations numerically in Sect.~\ref{sec:Solutions}.  
\section{$R^{2}$ Gravity Vacuum Equations} \label{sec:Quadratic EH}

\subsection{General equations}

The general Euler-Lagrange equations that arise from the Lagrangian in Eq.~(\ref{eq:LagrangianDensity}) have the following well-known form

\begin{equation} \label{eq:EFE}
R_{\mu \nu} - \frac{1}{2} g_{\mu \nu} R + \frac{2}{m^{2}} \left( \left(R_{\mu \nu} - \frac{1}{4} g_{\mu \nu} R \right) R    +  \nabla_{\mu} \nabla_{\nu} R - g_{\mu \nu} g^{\rho \sigma}  \nabla_{\rho} \nabla_{\sigma}R \right ) = 0
\end{equation}
where $\nabla_\mu$ is the covariant derivative and the Ricci tensor is defined 
as the contraction on the first and fourth indices of the Riemann tensor\footnote{In comparison with \cite{KMP1987}, our definition of the Ricci tensor is opposite in sign, as well as our $m^2 = 6 M^2$ of \cite{KMP1987}}:
\begin{equation} \label{eq:Riccitendef}
R_{\mu \nu} = R^{\lambda}_{\mu \nu \lambda} = \partial_{\nu} \Gamma^{\lambda}_{\mu \lambda} - \partial_{\lambda} \Gamma^{\lambda}_{\mu \nu} +  \Gamma^{\rho}_{\mu \lambda} \Gamma^{\lambda}_{\rho \nu} - \Gamma^{\rho}_{\mu \nu} \Gamma^{\lambda}_{\rho \lambda}
\end{equation}
The contracted or trace equation will also be useful later:
\begin{equation}\label{eq:TraceEFE}
g^{\mu \nu}\nabla_{\mu}\nabla_{\nu} R + \frac{m^{2}}{6}R = 0  
\end{equation}

Overall, equations of $R^2$ gravity involve second derivatives of the Ricci tensor and, therefore,
fourth derivatives of the metric coefficients.
It is also clearly seen that $R_{\mu \nu} = 0$ is a solution to these equations, meaning that any Ricci-flat space-time is an exact solution for $R^2$ gravity.

\subsection{$R^{2}$ Gravity With Homogeneous and Isotropic Flat Metric} \label{sec:IsotropicGravity}

Let us briefly review how $R^{2}$ gravity works in case of homogeneous and isotropic 
space-times with flat spatial sections.  This will serve as a benchmark and introduction the notations that we will use for anisotropic space-times in the next section.

The Friedmann-Robertson-Walker (FRW) homogeneous and isotropic metric is given by:

\begin{equation}
    ds^{2} = dt^{2} - a^{2}(t)[dx^{2} + dy^{2} + dz^{2}]
\end{equation}
where $a(t)$ is the scale factor. 

The non-zero Ricci tensor components and the Ricci scalar can be expressed concisely
via the Hubble parameter $H=\frac{\dot{a}}{a}=\dot{(\ln a)}$ as
\begin{equation}\label{eq:RinFRW}
\begin{aligned}
& R_0^0 = 3 (\dot{H}+H^{2}), \quad
    R^i_J = (\dot{H}+3H^{2})\delta^i_j,   \quad
    R = 6 (\dot{H}+ 2 H^{2}), \quad
    \dot{R} = 6 (\ddot{H}+4H\dot{H})
\end{aligned}
\end{equation}
where Latin indices denote the 3-dimensional spatial coordinates, and Greek indices the 4-dimensional spacetime. 

As the equations for the isotropic universe we write down the $00$ component of Eq.~\ref{eq:EFE}
that reduces to the quadratic equation for the Hubble parameter
\begin{equation} \label{eq:FriedmannRsquared}
m^{2} H^{2} - 6\dot{H}^{2} + 36H^{2}\dot{H} + 12\ddot{H}H = 0
\end{equation}
This equation allows for a phase space analysis in $(H,\dot H)$ plane.  Such analysis was
performed (as part of a more general setup with conformal anomaly) for the relevant to cosmology expanding solutions in \cite{KMP1987}.  The most striking property is the existence of the intermediate
attractor along the separatrix that follows $\ddot{H} \approx 0, \dot{H} \approx - \frac{m^2}{36}$ line for $ |\dot{H}| \ll H^2$
and then winds up in an oscillatory spiral in the vicinity of $H=0, \dot H=0$ singular point (see Fig.~2 of \cite{KMP1987})
This is, of course, a famous Starobinsky \cite{Starobinsky} inflationary solution 
\begin{equation}\label{eq:Htslowroll}
H \approx H_0 - \frac{1}{36} m^2 t ~,
\end{equation}
that gracefully exits into the dust-like ``scalaron'' oscillations when the slow-roll condition  $|\dot{H}| < H^2$ breaks down.

However, the Eq.~\ref{eq:FriedmannRsquared} does not allow to invesitgate the system in all the details.
Indeed, the only finite singular point, $H=0,\dot{H}=0$ in the phase diagram, is \textit{not} a stationary point.
The system can pass through it, following multiple possible trajectories that intersect at $H=0,\dot{H}=0$, including the spiralling part
of the attracting separatrix (associated with a stationary point at infinity).
In particular, this equation cannot be used for numerical studies.  The reason is obvious. 
$(00)$ Einstein equation that was used is a contraint equation that describes the integral of motion
and has less degrees of freedom (three), that what the whole system does (four) (one additional silent degree of freedom is in going from Hubble parameter as a variable to the scale factor).

For numerical studies we need to use the equation with the full range of freedom, for instance the trace
of Einstein equations Eq.~\ref{eq:TraceEFE} that becomes
\begin{equation}\label{eq:traceiso}
\ddot{R} + 3 H \dot R + \frac{m^2}{6} R =0
\end{equation}

Let us introduce the following set of variables that will also be used for anisotropic studies
\begin{equation}{\label{eq:variables}}
y = 3H = \frac{d}{dt} \log(a^3),\quad z=\dot{y}, \quad \omega=\dot{R}
\end{equation}
which describes the change in the comoving volume of the space through $y$,  and the rate of change of  4-curvature $R$.
The scale factor and Ricci curvature $R$ can be reconstructed from these parameters as
\begin{equation}
a = a_0 e^{\frac{1}{3} \int y dt},    \quad   R = 2 z + \frac{4}{3} y^2 
\end{equation}

Using the expression for $\dot{R}$ from Eq.~\ref{eq:RinFRW} we can reduce the trace equation Eq.~\ref{eq:traceiso}
to three equations of the first order
\begin{equation} \label{eq:IsotropicEQs}
\begin{aligned}
& \dot{w} = -yw - \frac{m^{2}}{9} \left(2 y^2 + 3 z\right) \\
& \dot{y} = {z} \\ 
& \dot{z} = \frac{1}{2} w - \frac{4}{3}yz
\end{aligned}
\end{equation}
that describe the full set of dynamical degrees of freedom.
These equations will be useful for comparison with the anisotropic case. We also have the constraint equation, where Eq.~\ref{eq:FriedmannRsquared} becomes:
\begin{equation} \label{eq:xyFriedmannRsquared}
m^{2} y^{2} - 6 z^{2} -4 y^{2} z + 6 y w  = 0
\end{equation}
that must be satisfied by the initial conditions.

\section{Derivation of $R^{2}$ Anisotropic Gravity Equations} \label{sec:AnisotropicGravity}

We now turn our attention to
the anisotropic Kantowski-Sachs space \cite{Kantowski}. The Kantowski-Sachs metric  is given by

\begin{equation} \label{eq:KanSach}
ds^{2} = dt^{2} - a^{2}(t)dx^{2} - b^{2}(t)[dy^{2} + dz^{2}]
\end{equation}
where $a(t)$ and $b(t)$ are two unequal scale factors.

Computing the Ricci tensor components from the definition of the Ricci tensor given in Eq.~\ref{eq:Riccitendef} one obtains the following non-zero components

\begin{equation} \label{eq:KantowskiRicci}
\quad R_0^0 = \frac{\ddot{a}}{a} +  2\frac{\ddot{b}}{b}, \quad R_1^1 = \frac{\ddot{a}}{a} + 2\frac{\dot{a}\dot{b}}{ab}, \quad R_i^j = \left(\frac{\ddot{b}}{b} + \frac{\dot{b}^{2}}{b^{2}}  + \frac{\dot{a}\dot{b}}{ab}\right)\delta_i^j
\end{equation}
and the Ricci scalar
\begin{equation}
R =  2 \frac{\ddot{a}}{a} +  4 \frac{\ddot{b}}{b} + 2\frac{\dot{b}^{2}}{b^{2}} + 4\frac{\dot{a}\dot{b}}{ab}
\end{equation}
In this section Latin indices $i,j$ vary only between $2,3$, while the first spatial direction is
treated individually.

\subsection{x, y, z, w Equations}
Analogously to the $R^{2}$ isotropic gravity case, it is helpful to define two new variables $x = \frac{d}{dt} {\log(\frac{b}{a})}$ and $y  = \frac{d}{dt} {\log(ab^{2})}$. The variable $y$ is again a measure of the comoving volume rate of change, but a new variable $x$ is a measure of the anisotropy.
Expressing the non-zero components of Ricci tensor in terms of these variables
\begin{equation} \label{eq:xyRicci}
R_{0}^{0} = \frac{2}{3}x^{2} + \frac{1}{3}y^{2} + \dot{y}, ~~ R_{1}^{1} = -\frac{1}{3} \left(y^{2} - xy + 2\dot{x} - \dot{y}\right), ~~ R_{i}^{j} = \frac{1}{3}\left(xy + y^{2} + \dot{x} + \dot{y}\right) \delta_{i}^{j}
\end{equation}
one obtains the Ricci scalar and its derivative as
\begin{equation}\label{eq:xyRicciScalar}
\quad R = \frac{2}{3}x^{2} + \frac{4}{3}y^{2} + 2\dot{y}, \quad \dot{R} = \frac{4}{3}\dot{x}{x} + \frac{8}{3}\dot{y}y + 2\ddot{y}
\end{equation}
Retaining the definitions $z=\dot{y}$ and $\omega = \dot{R}$, after lengthy algebraic manipulations with
the help of symbolic algebra software, we obtain a set of four equations of motion
\begin{equation} \label{eq:5ODEs}
\begin{aligned}
&\dot{w} = -yw - \frac{m^{2}}{9} (x^2 + 2 y^2 + 3 z ) \\ 
&\dot{x} = -x y -  \frac{6 x w}{3m^{2} + 4x^{2} + 8y^{2} + 12{z}} \\ 
&\dot{y} = {z} \\ 
&\dot{z} =  \frac{1}{2} w  - \frac{4}{3} y z + \frac{2}{3} x^2 y + \frac{4 x^2 w }{3m^{2} + 4x^{2} + 8y^{2} + 12{z}}
\end{aligned}
\end{equation}
subject to the constraint from the (0,0) component of Eq.~\ref{eq:EFE}
\begin{equation} \label{eq:xy00Eq}
\begin{aligned}
& m^{2}(y^{2} - x^{2}) +  6 y \omega -6 z^2 - 4 z (y^2+2 x^2) -2 x^2 (x^2+2y^2) = 0 
\end{aligned}
\end{equation}
It can be seen that if we set $x = 0$, we recover the isotropic case, Eq.~\ref{eq:IsotropicEQs} and Eq.~\ref{eq:xyFriedmannRsquared}. Moreover, if one starts from the state with no initial anisotropy, the anisotropy will not be generated during the evolution, since $\dot{x} = 0$. 
In general, Kantowski-Sachs metric has two additional degrees of freedom in $R^2$ gravity
as compared to the isotropic case, to the total of six.  One extra degree of freedom is in 
the parameter $x$ while the second one is in recovering of two scale factors instead of one from $x$ and $y$.

\subsection{Kasner solution}
At the focus of this paper is Kasner \cite{kasner} solution in the class of Kantowski-Sachs metrics \cite{Kantowski}.  Kasner solution describe anisotropic expansion of a spatially flat space-time
and is a vacuum solution of Einstein equations. As such, it is Ricci-flat,  $R_{\alpha\beta}=0$,
and remains the exact solution with the addition of the $R^2$ term in the Lagrangian as well.
In fact, this extends to general $f(R)$ gravity \cite{Shamir}.  In case of the metric in Eq.~\ref{eq:KanSach}, Kasner solution gives
\begin{equation}\label{eq:KasnerinKS}
a(t) = a_0 (t/t_0)^{-1/3}   \quad  b(t) = b_0 (t/t_0)^{2/3}
\end{equation}
which describes the expansion of the space-time in two directions and contraction in the remaining one
(assuming the increasing time variable). 
The comoving volume in such space is changing as $a b^2 \propto t/t_0$.

In the language of $(x,y,z,\omega)$ variables, Kasner solution corresponds to
\begin{equation}
x = \frac{1}{t}, \quad y=x=\frac{1}{t},  \quad z=-y^2=-\frac{1}{t^2},  \quad \omega = 0
\end{equation}
It can be easily seen that it satisfies the system of Eq.~\ref{eq:5ODEs} as
well as the constraint Eq.~\ref{eq:xy00Eq} and has $R=0$.
\section{Numerical Solutions and Stability of Kasner expansion} \label{sec:Solutions}
\subsection{Problem setup and initial conditions}

In Einstein gravity, there are no non-trivial homogeneous and isotropic flat solutions without the matter.  In particular,
Kasner solution with equal exponents along all three axes is not allowed.  Thus, initially anisotropic
space-times will remain anisotropic. With $R^2$ term in Lagrangian this changes.  Isotropic solutions in vacuum are possible and we can
ask whether initially anisotropic spaces will isotropize as they evolve. We investigate numerically whether this happens for expanding 
spaces that are nearly Kasner in the beginning, or whether the solutions remain close to Kasner-like behaviour.

We specify the initial conditions at time $t_0$ to be nearly Kasner, subject to small deviation.
We choose this perturbation to occur in the highest order derivative quantities, $z$ and $\omega$, to have maximally smooth deviation
from Kasner law. 
\begin{equation}\label{eq:InitialConditions}
x_{0} = \frac{1}{t_{0}}, \quad y_{0} = \frac{1}{t_{0}}, \quad z_{0} = -\frac{1}{{t_{0}}^{2}}(1+p),  \quad w_{0} = \frac{p^{2}}{{t_{0}}^{3}} 
\end{equation}
We parameterize the deviation by the dimensionless parameter $p$, such that $p \geq 0$ corresponds to solutions where the volume expansion slows down faster than in Kasner regime, while $p \leq 0$ corresponds to solutions where volume expansion is less decelerated.  Notice that the initial value of $\omega$ is determined by the (0,0) constraint, Eq.~\ref{eq:xy00Eq} and is not free.  This initial configuration is not Ricci-flat,  $R_{0} = - \frac{2p}{{t_{0}}^{2}}$ with the sign that is determined by $p$, while the initial curvature derivative $w_0$ is always positive. 
Thus, for positive $p$, initial evolution will tend compensate the deviations of Kasner $R=0$, while the negative $p$ will accelerate
this deviation.

\subsection{Bifurcation in initial conditions}
Even a cursory examination of Eqs~\ref{eq:5ODEs} shows that there is a singular situation when the initial conditions are chosen to satisfy
$3m^{2} + 4x^{2} + 8y^{2} + 12{z} = 0 $.  With our starting state given by Eq.~\ref{eq:InitialConditions}, this corresponds to
\begin{equation}
p = \frac{1}{4} m^2 t_0^2
\end{equation}
curve in the $(t_0,p)$ parameter space.  As we shall show numerically next,  this is a bifurcation line that separates two vastly
different outcomes for the model.  If started below the bifurcation line,  $p < \frac{1}{4} m^2 t_0^2$, including all negative $p$, 
(except $p=0$), the anisotropic near-Kasner solution will
reverse the contraction of one axis into its expansion of all three axes,  eventually leading to
isotropic expansion of space. While for $p > \frac{1}{4} m^2 t_0^2$, the collapsing first axis will accelerate its collapse,
eventually reversing the initially growing comoving volume into a collapsing one that reaches singularity in a finite time. 

Another way to formulate the bifurcation condition, is as a condition on the initial scalar curvature, $R_0 = -\frac{1}{2} m^2$.
The universe collapses, if the initial curvature value (zero on Kasner solution) is more negative than that.

\subsection{Results}

Now we numerically solve our equations.  We can absorb the parameter $m$ into definition of time $m t \to t$, by setting $m=1$.
This assumes that quantum corrections are always present, and significant for $ t < 1 $, while having diminishing contribution for $t > 1$.
Lastly, we choose $a_{0} = -\frac{1}{3}\log{t_{0}}$ and $b_{0} = \frac{2}{3}\log{t_{0}}$ for the scale factors, 
which for purely Kasner solution gives $a=b=1$ for $t=1$.

\begin{figure}[ht]
\includegraphics[width=10 cm]{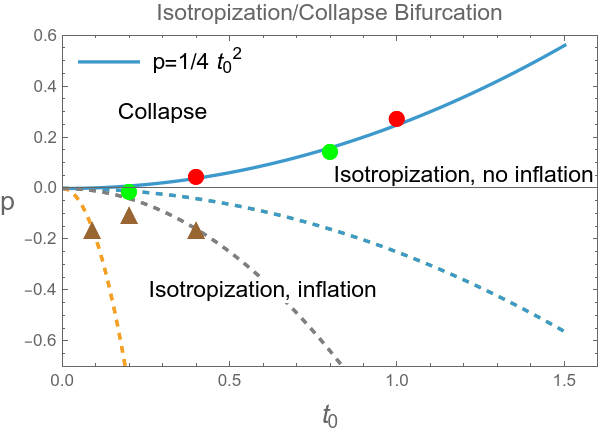}
\caption{The behaviour of the universe is determined by the values of $t_{0}$ and $p$ that are chosen. We found two clear situations, an isotropizing expanding universe and an anisotropically collapsing universe. The solid blue line indicates the bifurcation line between the regimes, where any point above that line would lead to a collapsing universe. Any point below that line would lead to an isotropizing universe, except along the 
line $p=0$ which gives the exact Kasner solution.  The solid circles denote models that straddle the bifurcation line in our numerical tests. The red circles are just above it, and the green are just below.
The dashed lines indicates the more quantitative assessment of isotropization roughly specifying the lines along each the amount
of inflation is the same. There is no discernible stage of exponential behaviour above $p=-\frac{1}{4} t^2$ blue dashed line.
To the left of the orange dotted line, the amount of inflation exceeds 65-efolds, the nominal value necessary for successful cosmological models.
Overplotted are markers which indicate the initial conditions of other figures in this paper.} 
\label{fig:BifurcationLine}
\end{figure}

As stated earlier, we will consider a universe with an initially expanding volume, which corresponds to $y > 0$, or iterating forward in time from $t_{0}$.  We shall consider a set of models with parameters summarized in Fig.~\ref{fig:BifurcationLine}. 

\begin{figure}[ht]
\includegraphics[width=0.47\textwidth]{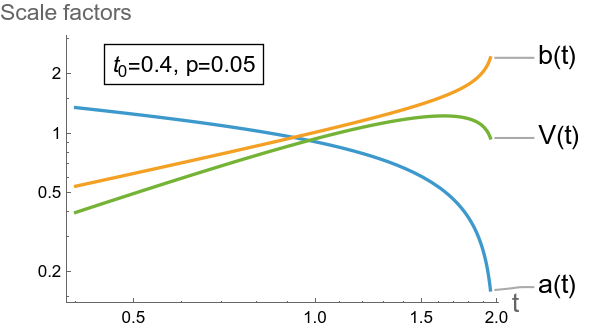} \hfill
\includegraphics[width=0.47\textwidth]{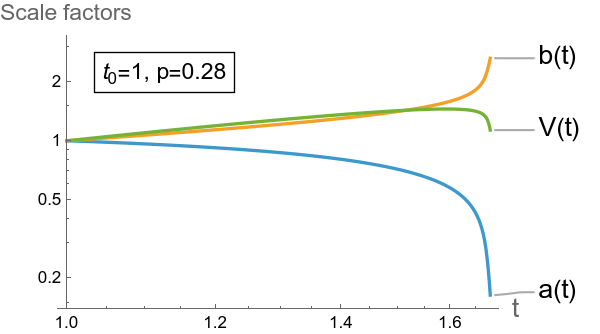}\\
\includegraphics[width=0.47\textwidth]{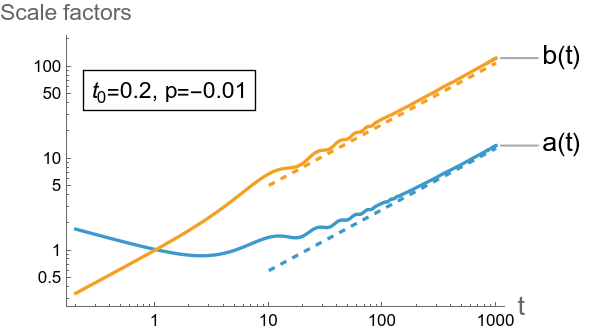} \hfill
\includegraphics[width=0.47\textwidth]{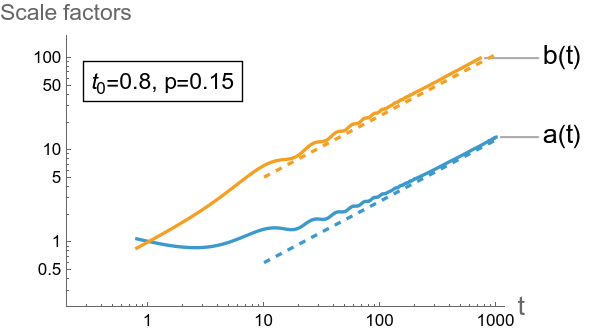}
\caption{A set of solutions with initial conditions that straddle the bifurcation line $p=\frac{1}{4} t_0^2$ corresponding to closed
dots in Fig.~\ref{fig:BifurcationLine}. Above the bifurcation line (top row) the space experiences fast accelerated anisotropic collapse,
with reversal of the initial volume expansion to contraction. Below the bifurcation line (bottom row) initial anisotropic Kasner-like expansion
transfers to isotropic one, following the ``dust-like" behaviour $a,b \propto t^{2/3}$ with superimposed decaying oscillations. The dashed lines show
$t^{2/3}$ asymptotics.}
\label{fig:IsotropizationSolution}
\end{figure}

As the first step, in Fig.~\ref{fig:IsotropizationSolution} we demonstrate that the singular line $p=\frac{1}{4} t_0^2$ in initial conditions
is indeed the bifurcation line.
Four models plotted in Fig.~\ref{fig:IsotropizationSolution} were chosen to straddle this line, two just above and two just below.
The outcomes are discontinuously different.  $p  > \frac{1}{4} t_0$ models collapse, the first collapsing axis of Kasner solution accelerates its decay
and eventually leads to the volume to shrink to a singularity as well.  $p < \frac{1}{4} t_0$ models have a dramatically different outcome.  The collapse
of the first axis is reversed, and the models enter the epoch of asymptotic growth of equal rate along all axes,  following the isotropic solutions
of $R^2$ gravity.  Thus, initially anisotropic, such spaces isotropize as $ t \to \infty $.  For two models of this type shown in Fig.~\ref{fig:IsotropizationSolution},
the isotropization happens shortly after $t=1$, at $t \approx 2$.  However, for $p$ closer to zero, the transition moment shifts to ever later times,
and, of course, for a special case of $p=0$ the solution remains strictly Kasner and anisotropic forever.

The models close to the bifurcation line, as in the bottom row of Fig.~\ref{fig:IsotropizationSolution}, shows isotropic transition practically 
straight into the "dust-like" expansion of the isotropic Universe with $a(t), b(t) \propto t^{2/3}$, with oscillations on top, which corresponds
to winding down spiral behaviour in $(H,\dot{H})$ phase space of Eq.~ref{eq:FriedmannRsquared} \cite{KMP1987}. Indeed, further numerical
studies show that this direct transition to ``dust-like'' limit is the outcome of all models in the sector $-\frac{1}{4}t_0^2 \lesssim p < \frac{1}{4} t_0^2$.
One caveat to note is that while the upper boundary of this range is a sharp, theoretically motivated, bifurcation line, the lower boundary comes from a more qualitative interpretation of numerical outcomes. 

\begin{figure} [h]
\includegraphics[width=0.47\textwidth]{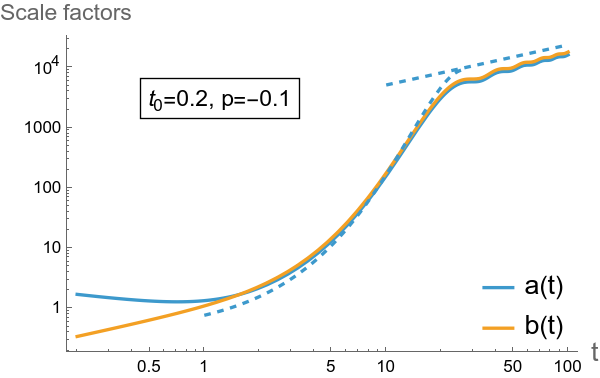} \hfill
\includegraphics[width=0.47\textwidth]{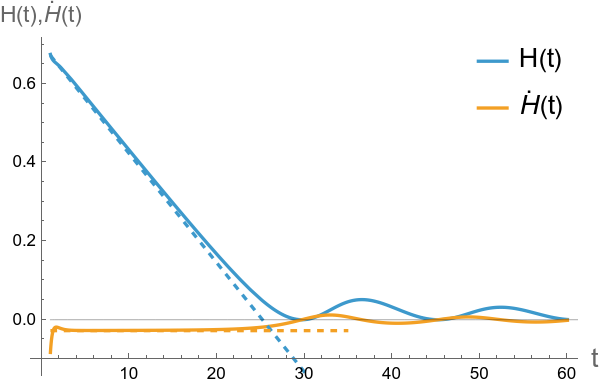}
\caption{A solution with the intermediate stage of near-exponential inflationary expansion. The left panel shows the scale factors
$a(t)$ and $b(t)$ which approach isotropy $a(t) \propto b(t)$ at $t > 2$.  The straight dashed lines corresponds to $\propto t^{2/3}$ 
``dust''-like asymptotitcs at $t \to \infty$. The rising dashed line is slow-roll Starobinsky inflationary regime of $R^2$ gravity 
$a(t) \sim e^{H_0 t - \frac{1}{72} m^2 t^2}$ that ends in this model at $ m t \approx 30$.  The right panel shows the behaviour of the Hubble
parameter and its derivative, $H(t)=\frac{1}{3}y(t)$, $\dot{H}(t) = \frac{1}{3} z(t)$. Inflationary stage is clearly seen, corresponding
to $\dot{H}(t) \approx -\frac{1}{36} m^2 = const$ and $H(t) = const -\frac{1}{36} m^2 t $, the behaviours shown by dashed lines.
On subsequent ``dust-like'' stage at $m t > 30$, $H$ and $\dot{H}$ oscillate, passing through $(0,0)$ point at every oscillation \cite{KMP1987}}.
\label{fig:InflationSolution}
\end{figure}

Next, Fig.~\ref{fig:InflationSolution} shows that for $p < -\frac{1}{4}t_0^2$, the isotropization occurs before entering the ``dust-like'' regime
and such model experiences a transition period of near-exponential isotropic expansion that is equivalent to Starobinsky inflation in $R^2$ gravity.
Such stage requires $| \dot{H} | \ll H^2$ and is described by Eq.~\ref{eq:Htslowroll} which in  our variables is $y(t) = y_0 - \frac{1}{12} t$.
The right panel of Fig.~\ref{fig:IsotropizationSolution} shows the presence of such slow-rolling change of the Hubble parameter from $t=2$ to $t \approx 30$.
Thus, the case investigated in this Figure has an intermediate inflationary asymptotics, which leads
to approximately 8 e-fold ($e^8$) expansion of the scale factors.  Note that the dimensional value of the Hubble expansion rate in our units at
is $H=\frac{1}{3} y \approx 0.7$ at the beginning of the isotropic regime, which is transitioned to at $t \approx 2$ again.

\begin{figure} [h]
\includegraphics[width=0.47\textwidth]{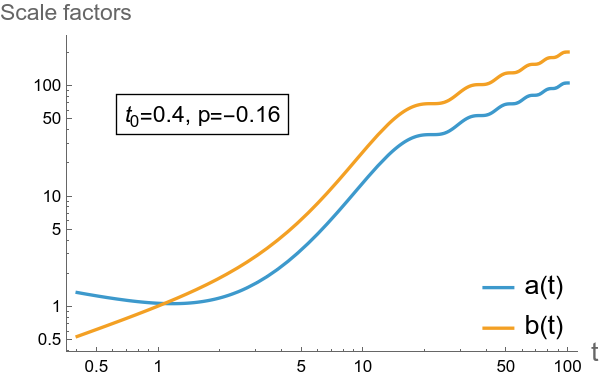} \hfill
\includegraphics[width=0.47\textwidth]{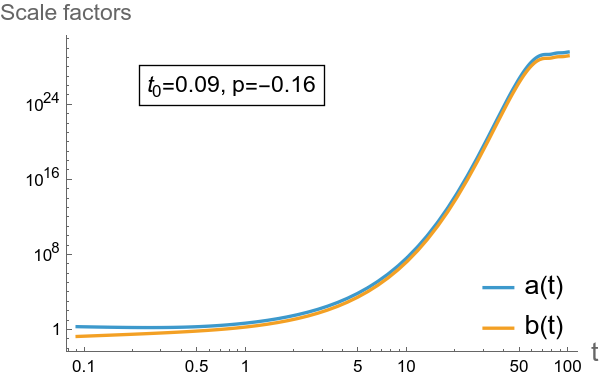}
\caption{Development of the inflationary stage for the parameters below $p < -\frac{1}{4} t_0^2$ line.  Left panel shows the example
of a model starting on $p=-t_0^2$ line (grey dashed line). These models show already a discernible, but very short 
inflationary stage, giving only 3-4 efold expansion.  The right panel shows example of the models that achieve cosmologically significant
benchmark expansion of 65 e-folds. Such models lie,
approximately, along the $p \approx -20 t_0^2$ line shown in Fig.~\ref{fig:BifurcationLine}.
}
\label{fig:ExponentialIsotropization}
\end{figure}

In Fig.~\ref{fig:ExponentialIsotropization}  we look at the
amount of exponential expansion that one obtains for different starting parameters.  We find that the models lying along the lines $p = - A t_0^2$,
$A > 1/4 $,  give approximately  the same number of e-folds during the inflationary stage.  Left panel of Fig.~\ref{fig:ExponentialIsotropization}
gives an example of a model with $A=1$ which has a noticeable, but very short inflationary stage. Let as take the nominal value
of 65 efolds as the required amount of inflation for a viable cosmological model. Such amount of expansion is achieved along $p=-20 t_0^2$ line
with the example case shown in the right panel of Fig.~\ref{fig:ExponentialIsotropization}. Note that all models which have an inflationary stage,
have transitioned to the isotropic stage at $t \approx 2$.  What is different between the model and what governs the duration of inflation, is
the value of the Hubble parameter $H=\frac{1}{3} y$ that transpires after isotropization. One needs $H/m \approx 1.9 $ to reach 65 e-folds.

\section{Conclusion} \label{sec:Conclusion}
We found that in $R^2$ gravity without matter fields, the anisotropic Kasner solutions,
are unstable and can evolve either towards isotropization of the expansion or collapse of the Universe.
What matters is how strong and how early is the deviation from the Kasner solution.
In fact, there is a clear bifurcation line in $p-t_0$ space, $p=\frac{1}{4} t_0^2$, as shown in Fig.~\ref{fig:BifurcationLine}, that
separates these two outcomes. Above this bifurcation line the universe will collapse anisotropically. Below the bifurcation line, the universe will continue to expand, and, in general, isotropize.
The condition can be formulated in terms of the initial 
perturbation in the scalar curvature, where  $R_0 < -1/2 m^2$ (using the choice of the curvature sign adopted in this paper) will result in the collapse and $R_0 > -1/2 m^2$ in the isotropization (except $R_0=0$).

It is interesting to consider the work of Toporensky and Müller once again \cite{Toporensky}. As they moved toward the BKL singularity, they found a bifurcation line between Kasner-type solutions and other solutions, but indicated that the Kasner-type solutions were much more stable than in our case of
starting with the Universe with expanding spatial volume, moving away from the BKL singularity. Forward-in-time  Kasner solution is shown to be unstable in this work, resulting either in a quick reversal of the volume expansion and collapse, or transition to isotropic expansion, but no solutions that continue to expand anisotropically except for the exact Kasner solution.

As we know, our universe today is expanding isotropically, and it is commonly accepted that there was an inflationary epoch shortly after the Big Bang.  We found that in $R^2$ gravity if the Universe was expanding anisotropically early on, subsequent isotropization and transition to Starobinsky isotropic model is
a generic outcome. For a sufficiently early and significant perturbation from Kasner behaviour there will be a period of near exponential expansion, before continuing to power law expansion with $a,b \propto t^{\frac{2}{3}}$.  For $R_0 \gtrsim 40 m^2$ the amount of inflation is sufficient to satisfy the cosmological constraints.

\section*{Acknowledgements} \label{sec:acknowledgements}
The authors would like to thank Alexei Starobinsky for fruitful discussions at the start of this project.

\bibliography{refs.bib}

\begin{thebibliography}{10}

\bibitem{Barrow}
John~D. Barrow and Sigbj\o{}rn Hervik.
\newblock Evolution of universes in quadratic theories of gravity.
\newblock {\em Phys. Rev. D}, 74:124017, Dec 2006.

\bibitem{belinsky}
V.~A. {Belinskij}, I.~M. {Khalatnikov}, and E.~M. {Lifshits}.
\newblock {Oscillatory approach to a singular point in the relativistic
  cosmology.}
\newblock {\em Advances in Physics}, 19:525--573, January 1970.

\bibitem{Davies}
P.~C.~W. {Davies}, S.~A. {Fulling}, S.~M. {Christensen}, and T.~S. {Bunch}.
\newblock {Energy-momentum tensor of a massive scalar quantum field in a
  Robertson-Walker universe.}
\newblock {\em Annals of Physics}, 109:108--142, January 1977.

\bibitem{Fischetti}
M.~V. {Fischetti}, J.~B. {Hartle}, and B.~L. {Hu}.
\newblock {Quantum effects in the early universe. I. Influence of trace
  anomalies on homogeneous, isotropic, classical geometries}.
\newblock {\em \prd}, 20(8):1757--1771, October 1979.

\bibitem{Ginzburg}
V.~L. {Ginzburg}, D.~A. {Kirzhnits}, and A.~A. {Lyubushin}.
\newblock {The Role of Quantum Fluctuations of the Gravitational Field in
  General Relativity Theory and Cosmology}.
\newblock {\em Soviet Journal of Experimental and Theoretical Physics}, 33:242,
  January 1971.

\bibitem{Kantowski}
R.~{Kantowski} and R.~K. {Sachs}.
\newblock {Some Spatially Homogeneous Anisotropic Relativistic Cosmological
  Models}.
\newblock {\em Journal of Mathematical Physics}, 7:443--446, March 1966.

\bibitem{kasner}
Edward {Kasner}.
\newblock {Geometrical theorems on Einstein's cosmological equations}.
\newblock {\em American Journal of Mathematics}, 43(4):217--221, October 1921.

\bibitem{KMP1987}
L.~A. {Kofman}, V.~F. {Mukhanov}, and D.~Iu. {Pogosian}.
\newblock {The evolution of inhomogeneities in inflationary models in the
  theory of gravitation with higher derivatives}.
\newblock {\em Zhurnal Eksperimentalnoi i Teoreticheskoi Fiziki}, 93:769--783,
  September 1987.

\bibitem{Nojiri}
S.~Nojiri, S.D. Odintsov, V.K. Oikonomou, and Arkady~A. Popov.
\newblock Ghost-free f(r,g) gravity.
\newblock {\em Nuclear Physics B}, 973:115617, 2021.

\bibitem{Shamir}
M.~Farasat {Shamir}.
\newblock {Some Bianchi type cosmological models in f( R) gravity}.
\newblock {\em \apss}, 330(1):183--189, November 2010.

\bibitem{Starobinsky}
A.~A. {Starobinsky}.
\newblock {A new type of isotropic cosmological models without singularity}.
\newblock {\em Physics Letters B}, 91(1):99--102, March 1980.

\bibitem{Toporensky}
A.~{Toporensky} and D.~{M{\"u}ller}.
\newblock {On stability of the Kasner solution in quadratic gravity}.
\newblock {\em General Relativity and Gravitation}, 49(1):8, January 2017.

\end{thebibliography}

\end{document}